\documentclass[12pt,noshowpacs,nofootinbib,notitlepage,amsmath,amssymb]{revtex4-2}
\usepackage{setspace}
\usepackage[top=1in,bottom=1in,left=1in,right=1in]{geometry}
\usepackage{graphicx,color}
\usepackage[colorlinks=true,citecolor=blue,linkcolor=blue,urlcolor=blue]{hyperref}
\usepackage{enumerate}
\usepackage{array}
\usepackage{bm,bbm}

\begin{document}

\title{\color{blue}\Large UV/IR scale dependence in a\\four-derivative scalar field theory}

\author{Bob Holdom}
\email{bob.holdom@utoronto.ca}
\affiliation{Department of Physics, University of Toronto\\ Toronto, Ontario, Canada  M5S 1A7}

\begin{abstract}
An analog of the standard renormalization procedure may be useful for four-derivative theories,
and we illustrate it with the four-derivative scalar field
theory in the $m\to0$ limit. The renormalization scale $\mu$ appears in momentum-dependent
logarithms associated with both UV and IR divergences. At one loop, the on-shell scattering
amplitude satisfies an analog Callan--Symanzik equation involving the analog beta functions.
These functions are determined from the combined UV and IR pole parts of the 1PI functions.
\end{abstract}

\maketitle
\section{Introduction}

One-loop amplitudes of four-derivative field theories contain
logarithms of both the ultraviolet renormalization scale $\mu_{\rm UV}$ and the
ghost/infrared mass scale $m$. Consequently, the standard $\beta$ functions obtained
from the $1/\varepsilon$ poles do not reflect the full logarithmic momentum dependence of
on-shell scattering. The possibility that the additional, $m$-dependent
logarithms define a distinct, ``physical'' running of the couplings has been
pursued most actively for quadratic gravity. Buccio, Donoghue, Menezes, and
Percacci~\cite{Buccio:2024hys} derived such $\beta$ functions from
a quadratic effective action using the background field method.

Salvio, Strumia, and Vitti~\cite{Salvio:2025cmi} looked more closely at amplitudes
in the spin-0 sector of quadratic gravity. They point out
that off-shell quantities, such as the one-loop propagator and the quadratic effective
action, exhibit spurious, gauge- and field-parameterization-dependent IR
running. Even for on-shell amplitudes, they claim a type of
process dependence in the form of unusual coefficients of momentum-dependent logarithms.
They also suggest that these logarithms in the four-derivative theory may have more in common with the IR
structure of a two-derivative theory in two dimensions than with the standard IR
divergences of a two-derivative theory in four dimensions.
Buccio, de Brito, and Parente~\cite{Buccio:2025tci} similarly argue that
the IR divergences in four-derivative theories are not of the usual soft or
collinear type, and are thus not amenable to standard factorization and/or cancellation methods.
These authors compute the one-loop momentum running of curvature-squared operators in quadratic and
conformal gravity using the background field method and find it explicitly
gauge dependent. They trace this dependence to the
two-point function and conclude that the gauge-invariant quantities are the
on-shell $n$-point amplitudes with $n\ge 4$ and that their explicit computation
is necessary to settle the UV behavior of the theory.

We shall consider the renormalization-group structure in the more
controlled setting provided by the shift-invariant four-derivative scalar theory
\cite{Holdom2023,Holdom2024}. This theory has both quartic and cubic four-derivative
couplings, and it is the cubic coupling that is required to generate the $\ln(-s/m^2)$
dependence in the on-shell amplitude. The $\ln(-s/m^2)$ terms come
from diagrams distinct from those used to calculate the UV-dependent $\beta$ functions.
In addition, $\ln(-s/m^2)$ contributes a finite
imaginary part to the amplitude, just as $\ln(-s/\mu_{\rm UV}^2)$ does. This behavior
differs from that of standard Sudakov double logarithms, which imply a divergent imaginary contribution.

Rather than focus directly on the momentum-dependent logarithms, particularly given their
complicated structure in three- and four-point functions, we follow
the standard renormalization procedure by calculating divergences in the 1PI
three- and four-point functions together with those in the two-point function. We then introduce
an analog of this procedure by defining a renormalization scale sensitive to both UV and IR divergences.

\section{The four-derivative scalar theory}

The shift-invariant four-derivative scalar field theory is~\cite{Holdom2023,Holdom2024}
\begin{equation}
\mathcal{L} = \frac{1}{2}\partial_\mu\phi(\Box + m^2)\partial^\mu\phi
+ \lambda_3(\partial_\mu\phi\partial^\mu\phi)\Box\phi
+ \lambda_4(\partial_\mu\phi\partial^\mu\phi)^2.
\label{eq:Lagrangian}
\end{equation}
The field $\phi$ and the couplings $\lambda_3$ and $\lambda_4$ are dimensionless.
The standard $\beta$ functions, computed from the UV $1/\varepsilon$ poles in dimensional
regularization, are~\cite{Holdom2023},
\begin{align}
\beta_3 &= \frac{d\lambda_3}{d\ln\mu_{\rm UV}}
= -\frac{5}{4\pi^2}\lambda_3\!\left(\lambda_4 + \frac{3}{4}\lambda_3^2\right),
\label{eq:beta3std} \\
\beta_4 &= \frac{d\lambda_4}{d\ln\mu_{\rm UV}}
= -\frac{5}{4\pi^2}\lambda_4\!\left(\lambda_4 + \lambda_3^2\right).
\label{eq:beta4std}
\end{align}
The one-loop anomalous dimension of $\phi$ is obtained from the self-energy
diagram, which is free of infrared divergences. The result is~\cite{Holdom2023}
\begin{equation}
\gamma = \frac{5\lambda_3^2}{16\pi^2}.
\label{eq:gamma}
\end{equation}
The RG flow of the standard $\beta$ functions is shown in Fig.~3 of Ref.~\cite{Holdom2023} and in Fig.~1 of
Ref.~\cite{Holdom2024}. The latter figure shows a red curve $\lambda_4 = -\frac{1}{2}\lambda_3^2$
that marks the boundary between two qualitatively different types of flow. On this
curve and for $m\to0$ the Lagrangian becomes a perfect square,
$\mathcal{L} = -\frac{1}{2}(\Box\phi - \lambda_3(\partial\phi)^2)^2$,
a theory of interest on its own~\cite{BatemanTurok2026,Anderson:2026ilf}.
Both the perfect-square theory and the general theory have well-behaved differential cross sections
when defined in a certain way, but here we are more interested in the renormalization structure.

If $m$ is kept as a fixed physical mass scale, then these $\beta$ functions are appropriate.
Above the mass scale, momentum-dependent logarithms associated with the UV divergences are accompanied by
additional momentum-dependent logarithms involving $m^2$. The latter can become large when
the energy is large compared with $m$. In the high-energy limit, one may instead prefer a
renormalization scale that better reflects the high energies. The limit $m/E\to0$ can be expressed
as $m\to0$ at fixed $E$. We are interested in the IR divergences that appear in this limit. The
$m\to0$ limit is also special in a four-derivative theory because $m$ separates the two poles
in the propagator and thereby provides a calculational method. For this reason, we consistently refer
to the $m\to0$ theory rather than the $m=0$ theory.

In dimensional regularization, both UV and IR divergences can produce the same
local structure: $1/\epsilon+\ln(\mu^2)$ for some renormalization
scale $\mu$ (we use $1/\epsilon=2/(4-d)-\gamma_E+\ln(4\pi)$). For example, in a
scaleless integral, the two contributions are equal and opposite and sum to zero.
In general, one needs a probe beyond dimensional regularization, such
as a small mass to determine the actual $[1/\epsilon+\ln(\mu^2)]_{UV}$ and
$[1/\epsilon+\ln(\mu^2)]_{IR}$ contributions. Normally it is only the former that
is relevant to renormalization.

The fact that dimensional regularization naturally produces both $[1/\epsilon+\ln(\mu^2)]_{UV}$ and
$[1/\epsilon+\ln(\mu^2)]_{IR}$ contributions is convenient for the $m\to0$ four-derivative theory
because the $[1/\epsilon+\ln(\mu^2)]_{IR}$ contributions have a role to play.
The $\ln(\mu^2)|_{IR}$ term allows the momentum dependence represented
by the finite $\ln(-s/m^2)$ in the original theory to appear instead as $\ln(-s/\mu^2)$
in the $m\to0$ theory. The $1/\epsilon|_{IR}$ term provides the analog of a counterterm for an
IR divergence and thereby leads to an analog of a renormalization procedure.
Meanwhile, the $[1/\epsilon+\ln(\mu^2)]_{UV}$ contributions play their usual role.
We use $\mu$ to denote a common
renormalization scale that is sensitive to both UV and IR effects. Thus, variation with respect
to this scale yields not the standard $\beta$ functions but their analogs, which we denote by $\hat\beta$.

\begin{figure}[t]
\centering
\includegraphics[width=\textwidth]{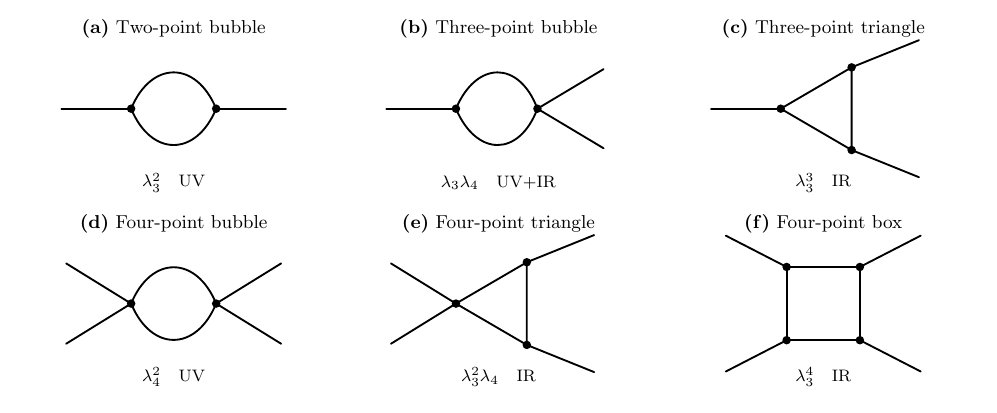}
\caption{The 1PI one-loop topologies relevant to the two-, three-, and four-point functions;
crossed external-leg configurations are not displayed.
The coupling factor and the origin of the pole are shown below each diagram.}
\label{fig:1pi-diagrams}
\end{figure}

In the analog renormalization procedure proposed here, both divergences can contribute to any
given analog counterterm. Although only the total contribution matters for each counterterm,
the relative $[1/\epsilon+\ln(\mu^2)]_{UV}$ and $[1/\epsilon+\ln(\mu^2)]_{IR}$ contributions from each diagram
can be determined.\footnote{For this and other calculations, we use Package-X~\cite{patel}.}
For the model at hand, the relevant 1PI topologies are collected in
Fig.~\ref{fig:1pi-diagrams}; these diagrams were originally discussed in~\cite{Holdom2023}.
For three-point diagrams, two of the three external lines are kept on shell, while for
four-point diagrams, all four external lines are kept on shell. The divergent local structures are
quartic in the external momenta, as required to renormalize a four-derivative theory.

Only the bubble-type diagrams, which have two propagators in the loop, produce UV divergences
and therefore enter the calculation of the standard $\beta$ functions. The mixed three-point
bubble in Fig.~\ref{fig:1pi-diagrams}(b) also has an IR divergence. The triangle and box
diagrams generate only IR divergences in the $m\to0$ limit. For the mixed three-point bubble,
the relative UV and IR contributions to the $1/\epsilon+\ln(\mu^2)$ structure are
\begin{align}
\left[\frac{1}{\epsilon}+\ln(\mu^2)\right]_{UV}+\left[\frac{1}{\epsilon}+\ln(\mu^2)\right]_{IR}=\left(\frac{15}{13}-\frac{2}{13}\right)\left(\frac{1}{\epsilon}+\ln(\mu^2)\right)
.\label{eq:split}
\end{align}

Analog renormalization can then be performed using the combined UV and IR divergences of the 1PI diagrams
to calculate the $\frac{1}{\epsilon}$ counterterms for the two couplings and the
wave-function renormalization of the four-derivative kinetic term. The associated
renormalization constants are given by (to compare with~\cite{Holdom2023}, we use $2/\varepsilon=1/\epsilon$)
\begin{align}
    (Z_\phi^\frac{3}{2}\hat Z_3-1)\lambda_{3}&=-\frac{13 \lambda _4 \lambda _3}{12 \pi ^2\varepsilon}-\frac{\lambda _3^3}{24 \pi ^2\varepsilon},\label{eq:Z3}\\
    (Z_\phi^2\hat Z_4-1)\lambda_{4}&=-\frac{5 \lambda _4^2}{4 \pi ^2\varepsilon}-\frac{7 \lambda _4 \lambda _3^2}{12 \pi ^2\varepsilon}-\frac{\lambda _3^4}{6 \pi ^2\varepsilon}.\label{eq:Z4}
\end{align}
$Z_\phi$ involves only a UV divergence and is unchanged. The first term in (\ref{eq:Z3}) changes because of the IR contribution in (\ref{eq:split}), while the first term in (\ref{eq:Z4}) is unchanged. The other three terms are new and arise from the three diagrams with only IR divergences.

The residues of the $\frac{1}{\varepsilon}$ poles in $\hat Z_{3}\lambda_{3}$ and $\hat Z_{4}\lambda_{4}$ give the analog $\hat\beta$ functions, for which $\ln\mu$ represents sensitivity to both UV and IR divergences,
\begin{align}
\hat\beta_3 &= \frac{d\lambda_3}{d\ln\mu}
= -\frac{13}{12\pi^2}\lambda_3\lambda_4 - \frac{47}{48\pi^2}\lambda_3^3,
\label{eq:beta3new} \\
\hat\beta_4 &= \frac{d\lambda_4}{d\ln\mu}
= -\frac{5}{4\pi^2}\lambda_4^2 - \frac{11}{6\pi^2}\lambda_3^2\lambda_4 - \frac{1}{6\pi^2}\lambda_3^4.
\label{eq:beta4new}
\end{align}
The anomalous dimension $\gamma$ remains as in (\ref{eq:gamma}). The second terms in (\ref{eq:beta3new}) and (\ref{eq:beta4new}) include the standard UV contributions from $Z_\phi$.

\begin{figure}[t]
\centering
\includegraphics[width=0.95\textwidth]{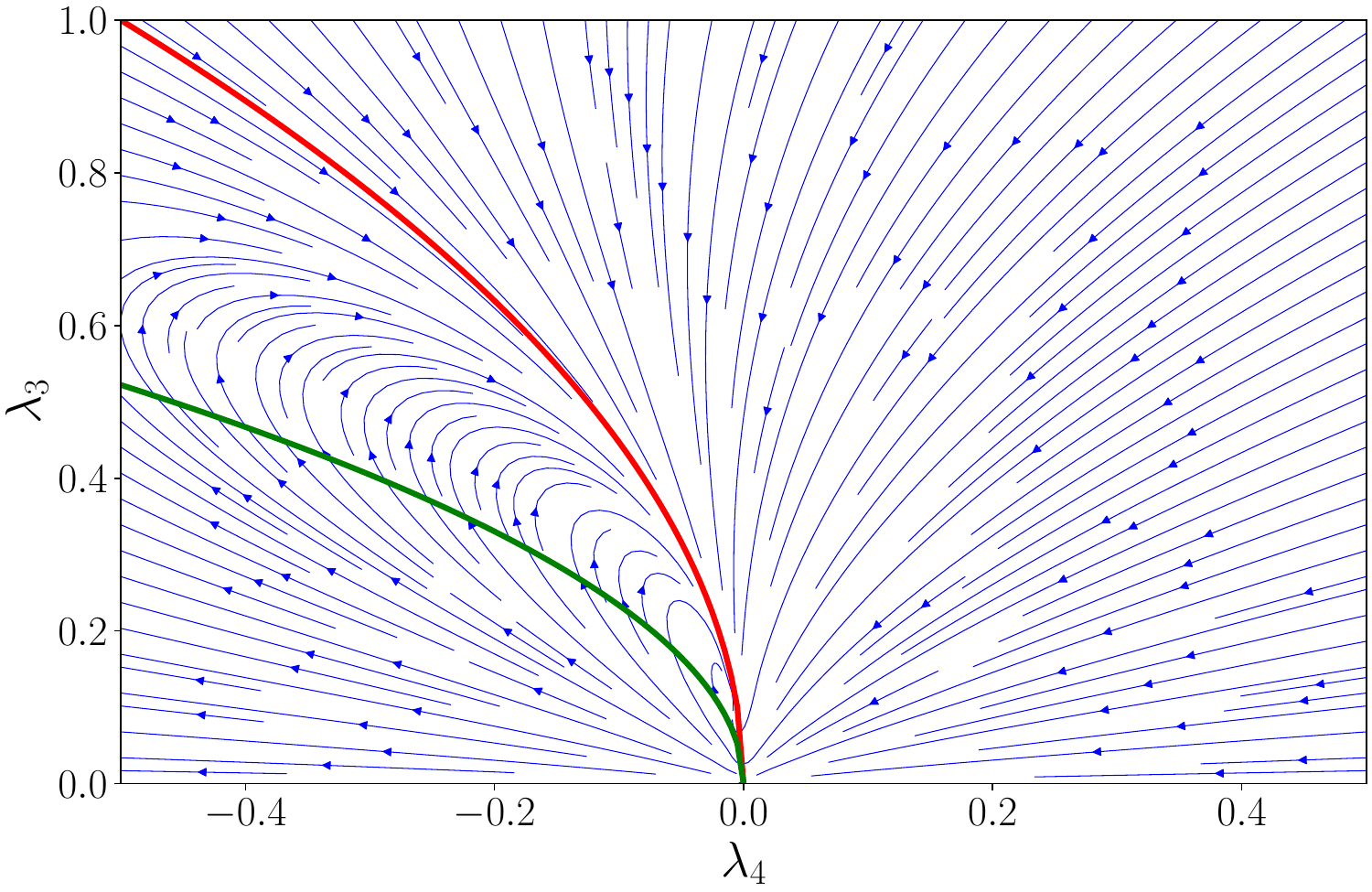}
\caption{RG flow generated by $\hat\beta_3$ and $\hat\beta_4$ in~\eqref{eq:beta3new} and~\eqref{eq:beta4new}.
The red curve $\lambda_4 = -\frac{1}{2}\lambda_3^2$ defines $D=0$ and has $dD/d\ln\mu=0$.
The green curve is another line where $dD/d\ln\mu=0$. When the flow lines cross the green curve they are
anti-parallel to the red flow line at the same $\lambda_3$.}
\label{fig:flow}
\end{figure}

The RG flow generated by the $\hat\beta$ functions is shown in Fig.~\ref{fig:flow}.
The curve $\lambda_4=-\frac{1}{2}\lambda_3^2$ plays a special role for the original $\beta$ RG flow
and continues to do so for the $\hat\beta$ RG flow. Defining
\begin{equation}
D = \lambda_4 + \frac{1}{2}\lambda_3^2
,\label{eq:Ddef}
\end{equation}
we see that this curve lies at $D=0$ and that $D$ is the horizontal displacement from it.
Its logarithmic derivative under the $\hat\beta$ RG flow is
\begin{align}
\frac{dD}{d\ln\mu} &= \hat\beta_4 + \lambda_3\hat\beta_3\nonumber\\
&=-\frac{5 \left(\lambda _3^2+2 \lambda _4\right) \left(11 \lambda _3^2+6 \lambda _4\right)}{48 \pi ^2}
.\label{eq:dDdt}
\end{align}
This derivative vanishes on the line $D=0$, which therefore remains a flow line. In addition, $dD/d\ln\mu \propto -D$ near that line, so the flow lines approach $D=0$ as $\mu$ increases, although they reach it only at $\lambda_3=\lambda_4=0$. There is also another line on which $dD/d\ln\mu=0$, namely $\lambda_4=-\frac{11}{6}\lambda_3^2$, but it is not a flow line.

The flow diagram for the standard $\beta$ functions shown in~\cite{Holdom2024} is qualitatively similar. The corresponding result is
\begin{align}
\frac{dD}{d\ln\mu_{\rm UV}} &= \beta_4 + \lambda_3\beta_3\nonumber\\
&=-\frac{5 \left(\lambda _3^2+2 \lambda _4\right) \left(3 \lambda _3^2+2 \lambda _4\right)}{16 \pi ^2}
.\label{eq:dDdt0}
\end{align}
The asymptotically free flow along the $D=0$ line satisfies $\frac{d\lambda_3}{d\ln\mu_{\rm UV}}=-5\lambda_3^3/16\pi^2$,
whereas $\hat\beta_3$ and $\hat\beta_4$ give $\frac{d\lambda_3}{d\ln\mu}=-7\lambda_3^3/16\pi^2$.

\section{Renormalization group equation}

The analog renormalized Lagrangian and its counterterms can be transformed into an analog
bare Lagrangian by absorbing the analog renormalization constants into the analog bare fields and couplings.
This construction motivates the analog Callan--Symanzik equation, which we now verify.

The standard Callan--Symanzik equation shows how the running of the couplings and the
field normalization compensate for a renormalized correlation function's explicit dependence
on the renormalization scale $\mu_{\rm UV}$. With the
definition of $\gamma$ from~\cite{Holdom2023}, the on-shell amputated connected four-point function $\Gamma^{(4)}$
satisfies the Callan--Symanzik equation
\begin{equation}
\left[\frac{\partial}{\partial\ln\mu_{\rm UV}}
+ \beta_3\frac{\partial}{\partial\lambda_3}
+ \beta_4\frac{\partial}{\partial\lambda_4}
+ 4\gamma\right]\Gamma^{(4)} = 0.
\label{eq:RG0}
\end{equation}
The scale derivative acts on the one-loop amplitude, while the other three terms involve
the tree-level amplitude. In the limit $m\to0$, these amplitudes depend on the Mandelstam variables
$s$, $t$, and $u$. The tree-level amplitude for the theory at hand is
\begin{equation}
\Gamma_{\rm tree}^{(4)}
= (\lambda_3^2 + 2\lambda_4)(s^2+t^2+u^2).
\label{eq:treeamp}
\end{equation}
The sum of the $\beta$-function terms in (\ref{eq:RG0}) is proportional to $dD/d\ln\mu_{\rm UV}$ in (\ref{eq:dDdt0}).

The analog Callan--Symanzik equation describes the explicit dependence on
the UV/IR renormalization scale $\mu$,
\begin{equation}
\left[\frac{\partial}{\partial\ln\mu}
+ \hat\beta_3\frac{\partial}{\partial\lambda_3}
+ \hat\beta_4\frac{\partial}{\partial\lambda_4}
+ 4\gamma\right]\Gamma^{(4)} = 0.
\label{eq:RG}
\end{equation}
It involves the analog $\hat\beta$ functions and the same $\gamma$. The sum of the $\hat\beta$-function terms
in (\ref{eq:RG}) is proportional to $dD/d\ln\mu$ in (\ref{eq:dDdt}).

\begin{figure}[t]
\centering
\includegraphics[width=\textwidth]{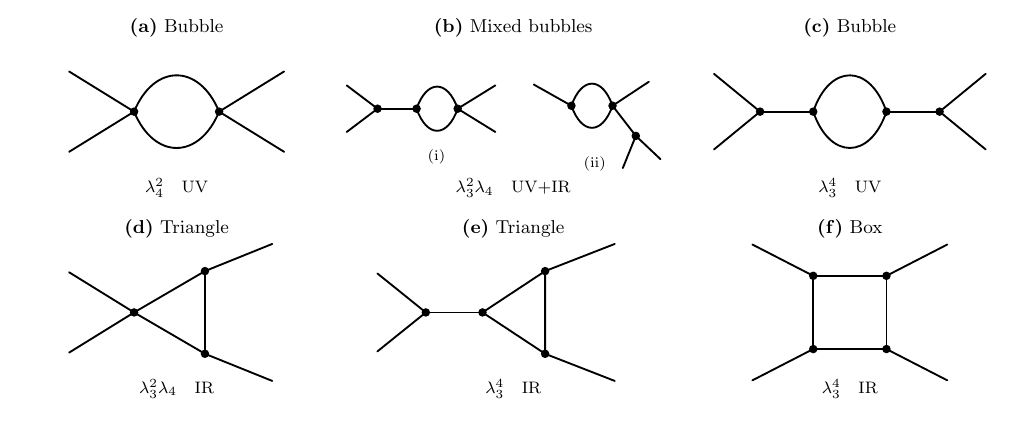}
\caption{The six classes of one-loop connected four-point diagrams. Panel (b) displays both
mixed-bubble attachments: the non-loop propagator connects two cubic vertices in (i), whereas
it connects the quartic vertex on the loop to a cubic vertex in (ii). Diagrams obtained by
crossing the external legs are not displayed.}
\label{fig:amplitude-diagrams}
\end{figure}

The calculation of the on-shell amputated connected one-loop amplitude $\Gamma_{\rm 1-loop}^{(4)}$ differs
from that of the renormalization constants from 1PI loops, particularly because the amplitude includes one-particle-reducible
contributions. We organize the contributions into the six classes shown in
Fig.~\ref{fig:amplitude-diagrams}. The labels ``bubble'', ``triangle'', and ``box'' correspond
to two, three, and four propagators in the loop, and each class is further distinguished by whether
it is proportional to $\lambda_4^2$, $\lambda_4\lambda_3^2$, or $\lambda_3^4$. These topologies
were used previously in the optical-theorem calculation of~\cite{Holdom2024}. In each case, we also
indicate whether the divergence is UV, IR, or both. A factor of $\frac{1}{16\pi^2}$ is implicit
throughout these results.

\begin{align}
\Gamma_{\rm bubble}^{\rm UV}|_{\lambda_4^2}&=\lambda_4^2 \left(s^2+t^2+u^2\right)\left(\frac{20 }{\epsilon
   }+\frac{56}{3}\right)\nonumber\\&+\frac{4}{3} \lambda_4^2 \left(13 s^2+t^2+u^2\right) \ln
   \left(-\frac{\mu ^2}{s}\right)+\frac{4}{3} \lambda_4^2
   \left(s^2+13 t^2+u^2\right) \ln \left(-\frac{\mu ^2}{t}\right)\nonumber\\&+\frac{4}{3}
   \lambda_4^2 \left(s^2+t^2+13 u^2\right) \ln \left(-\frac{\mu^2}{u}\right)
\end{align}
\begin{align}
\Gamma_{\rm bubble}^{\rm UV+IR}|_{\lambda_3^2\lambda_4}&=\lambda_3^2\lambda_4 \left(s^2+t^2+u^2\right)\left(\frac{52}{3\epsilon
   }+\frac{56}{9}\right)\nonumber\\&+\frac{52}{3} \lambda_3^2\lambda_4 \,s^2 \ln
   \left(-\frac{\mu ^2}{s}\right)+\frac{52}{3} \lambda_3^2\lambda_4
   \,t^2 \ln \left(-\frac{\mu ^2}{t}\right)+\frac{52}{3}
   \lambda_3^2\lambda_4 \,u^2 \ln \left(-\frac{\mu^2}{u}\right)
\end{align}
\begin{align}
\Gamma_{\rm bubble}^{\rm UV}|_{\lambda_3^4}&=\lambda_3^4 \left(s^2+t^2+u^2\right)\left(\frac{5}{\epsilon
   }+3\right)\nonumber\\&+5 \lambda_3^4 \,s^2 \ln
   \left(-\frac{\mu ^2}{s}\right)+5 \lambda_3^4
   \,t^2 \ln \left(-\frac{\mu ^2}{t}\right)+5
   \lambda_3^4 \,u^2 \ln \left(-\frac{\mu^2}{u}\right)
\end{align}
\begin{align}
\Gamma_{\rm triangle}^{\rm IR}|_{\lambda_3^2\lambda_4}&=\lambda_3^2\lambda_4 \left(s^2+t^2+u^2\right)\left(\frac{28}{3\epsilon
   }+\frac{236}{9}\right)\nonumber\\&+\frac{4}{3} \lambda_3^2\lambda_4 \left(5 s^2+t^2+u^2\right) \ln
   \left(-\frac{\mu ^2}{s}\right)+\frac{4}{3} \lambda_3^2\lambda_4
   \left(s^2+5 t^2+u^2\right) \ln \left(-\frac{\mu ^2}{t}\right)\nonumber\\&+\frac{4}{3}
   \lambda_3^2\lambda_4 \left(s^2+t^2+5 u^2\right) \ln \left(-\frac{\mu^2}{u}\right)
\end{align}
\begin{align}
\Gamma_{\rm triangle}^{\rm IR}|_{\lambda_3^4}&=\lambda_3^4 \left(s^2+t^2+u^2\right)\left(\frac{2}{3\epsilon
   }+\frac{10}{9}\right)\nonumber\\&+\frac{2}{3} \lambda_3^4 \,s^2 \ln
   \left(-\frac{\mu ^2}{s}\right)+\frac{2}{3} \lambda_3^4
   \,t^2 \ln \left(-\frac{\mu ^2}{t}\right)+\frac{2}{3}
   \lambda_3^4 \,u^2 \ln \left(-\frac{\mu^2}{u}\right)
\end{align}
\begin{align}
\Gamma_{\rm box}^{\rm IR}|_{\lambda_3^4}&=\lambda_3^4 \left(s^2+t^2+u^2\right)\left(\frac{8}{3\epsilon
   }+\frac{67}{9}\right)\nonumber\\&+\frac{1}{3} \lambda_3^4 \left(6 s^2+t^2+u^2\right) \ln
   \left(-\frac{\mu ^2}{s}\right)+\frac{1}{3} \lambda_3^4
   \left(s^2+6 t^2+u^2\right) \ln \left(-\frac{\mu ^2}{t}\right)\nonumber\\&+\frac{1}{3}
   \lambda_3^4 \left(s^2+t^2+6 u^2\right) \ln \left(-\frac{\mu^2}{u}\right)
\end{align}
$\Gamma_{\rm bubble}^{\rm UV}|_{\lambda_4^2}$ was obtained in~\cite{Tseytlin:2022flu,Holdom2023,Buccio:2023lzo}.
For $\Gamma_{\rm bubble}^{\rm UV+IR}|_{\lambda_3^2\lambda_4}$, the relative contributions of the two divergences to the coefficient of
$1/\epsilon+\ln(\mu^2)$ are given in (\ref{eq:split}).

The total unrenormalized one-loop amplitude is
\begin{align}
\Gamma_{\rm 1-loop}^{(4)}&=\left(\lambda _3^2+2 \lambda _4\right) \left(\frac{4}{9}(26 \lambda _3^2+21 \lambda _4)+\frac{5}{3}\frac{\left(5\lambda _3^2+6 \lambda _4\right) }{\epsilon }\right)\left(s^2+t^2+u^2\right)\nonumber\\&
+\frac{1}{3} \left(\lambda _3^2+2 \lambda _4\right) \ln \left(-\frac{\mu ^2}{s}\right)\left(\lambda _3^2 \left(23 s^2+t^2+u^2\right)+\lambda _4 \left(26 s^2+2 t^2+2 u^2\right)\right)\nonumber\\&
+\frac{1}{3} \left(\lambda _3^2+2 \lambda _4\right) \ln \left(-\frac{\mu ^2}{t}\right)\left(\lambda _3^2 \left(s^2+23 t^2+u^2\right)+\lambda _4 \left(2 s^2+26 t^2+2 u^2\right)\right)\nonumber\\&
+\frac{1}{3} \left(\lambda _3^2+2 \lambda _4\right) \ln \left(-\frac{\mu ^2}{u}\right)\left(\lambda _3^2 \left(s^2+t^2+23 u^2\right)+\lambda _4 \left(2 s^2+2 t^2+26 u^2\right)\right)
\label{eq:1loop}\end{align}
We see that $\Gamma_{\rm 1-loop}^{(4)}$ vanishes when $\lambda_4=-\frac{1}{2}\lambda_3^2$, just as $\Gamma_{\rm tree}^{(4)}$ does.

The analog renormalized version of $\Gamma_{\rm 1-loop}^{(4)}$ is obtained by using
the analog counterterm corresponding to the 1PI loop in each connected diagram to cancel
its $\frac{1}{\epsilon}$ pole.
We use this result to verify the analog Callan--Symanzik equation in (\ref{eq:RG}). Substituting $\Gamma_{\rm tree}^{(4)}$
and the results of the previous section into the final three terms in (\ref{eq:RG}) gives
\begin{align}
   \frac{\partial}{\partial\ln\mu}\Gamma_{\rm 1-loop}^{(4)}  -\frac{5 \left(\lambda _3^2+2 \lambda _4\right) \left(5 \lambda _3^2+6 \lambda _4\right)}{24 \pi ^2}\left(s^2+t^2+u^2\right)=0
.\end{align}
This equation is satisfied, as can be seen from the
$\frac{1}{\epsilon}$ coefficient in the unrenormalized result
(\ref{eq:1loop}), given that the divergence structure is $\frac{1}{\epsilon}+\ln(\mu^2)$
and that a factor of $\frac{1}{16\pi^2}$ is implicit.

The key point is that all the momentum-dependent logarithms in the analog renormalized one-loop amplitude involve the
renormalization scale $\mu$. The analog Callan--Symanzik equation shows that the tree-level result,
together with the $\mu$ dependence of the couplings and the anomalous dimension, determines the explicit $\mu$
dependence. This relation is the result of the UV/IR running induced by the $\hat\beta$ functions.

If we instead consider the standard renormalized version of $\Gamma_{\rm 1-loop}^{(4)}$, it
exhibits $\mu_{\rm UV}$ dependence that satisfies the standard Callan--Symanzik equation (\ref{eq:RG0}).
However, this version of $\Gamma_{\rm 1-loop}^{(4)}$ retains IR divergences.

The momentum dependence of the one-loop amplitude (\ref{eq:1loop}) is more intricate than its $\mu$
dependence because it is distributed nontrivially among the $\ln(s)$,
$\ln(t)$, and $\ln(u)$ terms. This is also true of the six separate contributions to the
one-loop amplitude, whether they are generated by UV or IR effects. Thus, each logarithm has
a coefficient that differs kinematically from the total $\ln(\mu^2)$ coefficient. This difference is
expected. Although the Callan--Symanzik equation, together with dimensional analysis, can be
used to determine the effect of a common dilation $(s,t,u)\to\rho(s,t,u)$, with ratios such as $t/s$
held fixed, it does not determine the dependence on those ratios.

The imaginary part of the one-loop amplitude is related even less directly to the $\ln(\mu^2)$ dependence,
in contrast to the close relation in ordinary theories between $\beta$ functions and the on-shell cuts
of diagrams. The imaginary part of the forward scattering amplitude was calculated in~\cite{Holdom2024} as
part of a check of the optical theorem. Although it still vanishes for $\lambda_4=-\frac{1}{2}\lambda_3^2$, the result
is distinctly different from the $\ln(\mu^2)$ coefficient,
\begin{equation}
2{\rm Im}\,\Gamma_{\rm 1-loop}^{(4)}(s,t\to0^-,u\to-s)
= \frac{s^2}{6\pi}(\lambda_3^2 + 2\lambda_4)(6\lambda_3^2 + 7\lambda_4).
\label{eq:optical}
\end{equation}
The imaginary parts of all six diagram classes, with UV and/or IR divergences, are needed for consistency
with the optical theorem. In effect, analog renormalization and the analog Callan--Symanzik equation provide
the same unified treatment of all six classes.

\section{Real emission}
In ordinary two-derivative theories, soft IR divergences in virtual corrections to
$2\to2$ scattering can cancel against divergences from the integration over
unresolved emission in  $2\to3$ scattering. We therefore consider the tree-level, on-shell
five-point amplitude in the four-derivative scalar theory. The connected,
amputated amplitude with $m\to0$ turns out to take the form of a simple generalization of $\Gamma_{\rm tree}^{(4)}$,
\begin{equation}
\Gamma_{\rm tree}^{(5)}=2\lambda_3 (\lambda_3^2+2\lambda_4)
              \sum_{1\leq i<j\leq 5}s_{ij}^2.
 \label{eq:a5}
\end{equation}
Here all momenta are outgoing and obey $k_i^2=0$ and $\sum_i k_i=0$, so that
\begin{equation}
 s_{ij}=2k_i\cdot k_j,\qquad\sum_{j\ne i}s_{ij}=0.
\end{equation}
For the $2\to3$ process, $k_1=-p_1$, $k_2=-p_2$, and
$k_i=p_i$ for $i=3,4,5$, where the $p_i$ have positive energies.

We first comment on what is already clear from this result. Since the five-point amplitude is polynomial in the invariants, it has no
soft or collinear poles. Unresolved single-particle emission therefore produces no soft or collinear
divergence to cancel the IR divergence of the one-loop four-point amplitude.
This confirms the unusual character of this IR divergence, as mentioned in the introduction,
and it provides further support for the idea that this
IR divergence requires its own analog of renormalization.

The five-point tree diagrams consist of one-propagator trees with one cubic and
one quartic vertex and two-propagator trees with three cubic vertices.
There are 10 of the former, obtained by choosing the two external legs at
the cubic vertex, and 15 of the latter, obtained by choosing the external leg
at the central vertex and one of the three pairings of the remaining four legs.
The propagator is $-i/(K^2+i0)^2$, and the vertices are $iV_n$, where, with all
vertex momenta outgoing,
\begin{align}
 V_3(a,b,c)
 &=2\lambda_3\bigl[(a\cdot b)c^2+(a\cdot c)b^2
                         +(b\cdot c)a^2\bigr],
 \label{eq:v3}\\
 V_4(a,b,c,d)
 &=8\lambda_4\bigl[(a\cdot b)(c\cdot d)
                  +(a\cdot c)(b\cdot d)
                  +(a\cdot d)(b\cdot c)\bigr].
 \label{eq:v4}
\end{align}
Both types of diagram have overall momentum degree
four, since a tree has one fewer internal line than vertices.

To understand the polynomial nature of $\Gamma_{\rm tree}^{(5)}$, consider a
cubic vertex connecting two external lines to an internal line, with
$a^2=b^2=0$ and $K=a+b$. Equation~\eqref{eq:v3} gives
\begin{equation}
 V_3(a,b,-K)=\lambda_3(K^2)^2,
 \label{eq:terminal}
\end{equation}
which cancels the denominator of the adjacent four-derivative propagator.
Every internal line of a five-point tree is adjacent to such a cubic vertex,
and thus each
diagram reduces to a polynomial.

We now examine the single-soft limit of $\Gamma_{\rm tree}^{(5)}$, which has already used the $m\to0$
limit. Write $k_5=\omega\ell$, where
$\ell=(1,\bm n)$ and $\bm n^2=1$, and let $\omega\to0$ along a smooth,
momentum-conserving trajectory toward generic four-point hard kinematics.
Set $r_i=s_{i5}$ for $i=1,\ldots,4$, with $\sum_{i=1}^4r_i=0$. Defining
\begin{equation}
 \bar s=\frac{s_{12}+s_{34}}2,\qquad
 \bar t=\frac{s_{13}+s_{24}}2,\qquad
 \bar u=\frac{s_{14}+s_{23}}2,
 \label{eq:averages}
\end{equation}
gives $\bar s+\bar t+\bar u=0$ including at nonzero $\omega$.
The $r_{i}$ generate deviations from $2\to2$ kinematics,
\begin{equation}
 s_{34}-s_{12}=r_1+r_2,\quad
 s_{24}-s_{13}=r_1+r_3,\quad
 s_{23}-s_{14}=r_1+r_4.
\end{equation}
One then finds the identity
\begin{equation}
\sum_{1\leq i<j\leq 5}s_{ij}^2=2(\bar s^2+\bar t^2+\bar u^2)
                  +\frac32\sum_{i=1}^4r_i^2,
\end{equation}
or equivalently
\begin{equation}
\Gamma_{\rm tree}^{(5)}
 =4\lambda_3\Gamma_{\rm tree}^{(4)}(\bar s,\bar t,\bar u)
       +3\lambda_3 (\lambda_3^2+2\lambda_4)\sum_{i=1}^4r_i^2.
 \label{eq:exact-averaged}
\end{equation}
Here the $\Gamma_{\rm tree}^{(4)}$ of~\eqref{eq:treeamp} is evaluated on the
averaged invariants. \eqref{eq:exact-averaged} implies the leading soft relation
$\Gamma_{\rm tree}^{(5)}\to4\lambda_3\Gamma_{\rm tree}^{(4)}(s,t,u)$, with an $O(\omega^0)$ soft factor rather than the usual $1/\omega$. The departure
from $4\lambda_3\Gamma_{\rm tree}^{(4)}(\bar s,\bar t,\bar u)$ starts at $O(\omega^2)$.
The averaged invariants themselves generally vary at $O(\omega)$, and this
linear correction relative to the limiting hard amplitude depends on the specific
recoil prescription.

One may wonder whether the polynomial form and the lack of soft and collinear divergences extend
to higher point amplitudes. At six points, propagator denominators adjacent to cubic vertices with two
massless external legs are again cancelled. However, an internal line can now
connect two nonzero four-point subamplitudes and carry the momentum
$K=k_1+k_2+k_3=-(k_4+k_5+k_6)$. The resulting $1/(K^2)^2$ pole survives in
the full amplitude, which is therefore not a polynomial. These poles do not,
however, imply a divergent single-soft limit at generic hard kinematics;
if one of the three momenta is $q\to0$, then
$K^2=(k_i+k_j+q)^2\to s_{ij}\ne0$. If all three momenta are hard with $k_i$ and $k_j$
becoming collinear then $K^2=s_{ij}+s_{ik}+s_{jk}\to s_{ik}+s_{jk}\ne0$, thus giving no collinear divergence.

\section{Quantum quadratic gravity (QQG)}
In Einstein quantum gravity, the one-loop virtual correction  and unresolved soft-graviton
emission are separately IR divergent. The divergences
cancel in an inclusive differential cross section that combines the
$2\to3$ real-emission contribution with the interference between the tree
and one-loop $2\to2$ amplitudes. The remaining soft correction depends on
the energy resolution, $\Delta E$, but not on the IR regulator~\cite{Weinberg:1965nx}.

QQG matches Einstein gravity for any tree-level scattering amplitude involving external massless gravitons only \cite{Dona:2015tra}.
If we consider $2\to3$ scattering in the high energy limit and take one graviton soft,
we should also consider the process with the soft graviton replaced by a soft spin-2 ghost.
Here soft means $m_2<\omega<\Delta E$ with $m_2\ll\Delta E$ and we are treating the ghost as an asymptotic state.
Summed over its five helicities, the ghost's polarization sum differs from the graviton's
only by a term proportional to the trace of the eikonal source, and this vanishes when the hard
particle is massless. Thus the coupling of the soft emitted ghost or graviton is the same, only the mass differs.
On the other hand, the ghost is a negative-norm state, so the
differential cross section for its real-emission carries a minus sign. The virtual correction
also involves a virtual ghost, and the residue of its propagator is also negative.
The massless and massive contributions to the virtual correction are bundled
together in the quartic propagator, and this is why the real ghost emission needs to be considered.

We first treat the massless and massive cases separately and add real and virtual
contributions for each case. Schematically, in the soft (eikonal) approximation,
\begin{equation}
\frac{d\sigma}{d\sigma_{\rm tree}}-1\simeq
\underbrace{B\left[\ln\frac{\Delta E}{m_g}+\ln\frac{m_g}{E}\right]+C}_{\rm graviton}
-\underbrace{\left(B\left[\ln\frac{\Delta E}{m_2}+\ln\frac{m_2}{E}\right]+C\right)}_{\rm ghost}
+O\!\left(\frac{m_2}{\Delta E}\right),
\label{eq:softcancel}
\end{equation}
where in each bracket the first term is from real emission and the second from the virtual correction.
$B$ is the usual coefficient determined by the eikonal soft-graviton factors of the hard particles~\cite{Weinberg:1965nx},
$C$ is a regulator-independent constant, and $E$ is the hard scale. The real--virtual cancellation removes the
IR-regulator $m_g$ dependence in the massless correction and the $m_2$ dependence in the massive one.
The remainder in each case is the same as the Einstein gravity result, $B\ln(\Delta E/E)+C$,
except that one has the opposite sign. We find that the full GR-like soft correction
vanishes up to $O(m_2/\Delta E)$. This example gives a first look at the fate of the GR-like
IR structure in QQG; of course if we consider soft energies below $m_2$ then the GR structure reappears.

However, it is argued that QQG generates additional
IR divergences at one loop as $m_2\to0$ due to its four-derivative structure~\cite{Buccio:2024hys,Salvio:2025cmi,Buccio:2025tci}.
Ref.~\cite{Salvio:2025cmi} finds such effects in on-shell scattering
amplitudes in the scalar sector of QQG, which has similarities to the
four-derivative scalar theory considered here.
For the latter theory we have found that one-loop on-shell amplitudes can contain an IR divergence while
the corresponding tree-level single-soft emission amplitude
does not. This novel feature motivates investigating whether analog renormalization and analog
$\beta$ functions can properly organize the IR-sensitive logarithms of on-shell QQG amplitudes.

\section{Conclusion}

We have described an analog of renormalization for the $m\to0$ limit of a
shift-invariant four-derivative scalar theory. The combined UV and IR pole parts
of the 1PI two-, three-, and four-point functions define the analog $\hat\beta$
functions. At one loop, the $\hat\beta$ functions, together with the usual anomalous
dimension, account for the complete dependence of the on-shell four-point
amplitude on the common UV/IR scale $\mu$. In particular, the analog
Callan--Symanzik equation combines the bubble, triangle, and box contributions,
even though their divergences have different origins. The resulting flow also
preserves the special line $D=\lambda_4+\frac{1}{2}\lambda_3^2=0$.
Both the tree and one-loop four-point amplitudes vanish on this line, and the perfect-square
theory remains asymptotically free, although $\lambda_3$ runs somewhat faster than
it does under the standard UV $\beta$ functions.

The $\hat\beta$ functions and the anomalous dimension determine the coefficient of the total
$\ln(\mu^2)$ dependence of the amplitude, but not the dependence on ratios such as $t/s$
or the absorptive part. The latter contains
independent information about the on-shell intermediate states, as is clear
from the optical theorem result. Thus, the analog running provides a compact
organization of logarithms that become large as $m/E\to0$ and suggests a
possible framework for resumming them, but it does not replace the calculation
of the complete amplitude.

Beyond one loop, overlapping UV and IR regions will
test whether their combined local pole parts continue to define a useful closed
evolution for on-shell amplitudes. The line $D=0$ offers another test;
its perfect-square form is protected under standard UV renormalization by an
all-order structure~\cite{Anderson:2026ilf}, and it remains to be understood
whether a corresponding principle controls the combined UV/IR evolution
beyond one loop.

We note that the fixed-scattering-angle behavior $\Gamma^{(4)}\sim s^2$
is not reflected in the two-to-two differential cross section discussed
in~\cite{Holdom2024,BatemanTurok2026,Anderson:2026ilf}, which at tree level behaves as
$d\sigma/d\Omega\sim1/s$ in the same $m\to0$ limit. This differential cross section can be obtained
either through an inclusive sum over the mass-eigenstate channels, or
by taking appropriate mass derivatives, before taking $m\to0$. It is also expressed as a certain covariant
external-state prescription in~\cite{BatemanTurok2026,Anderson:2026ilf}.
The tree level inclusive differential cross section is nonzero even on the $D=0$ line.
The loop correction to the $m\to0$ inclusive
differential cross section remains to be calculated, but obtaining a renormalized
result that is both UV and IR finite is now possible.

\section*{Acknowledgements}
I thank Jing Ren for prompting me to look again at IR divergences in four-derivative theories
and Yannick Kluth for prompting me to look at the 5-point function.

\end{document}